\documentclass[conference]{IEEEtran}
\IEEEoverridecommandlockouts
\usepackage{cite}
\usepackage{amsmath,amssymb,amsfonts}
\usepackage{algorithmic}
\usepackage{graphicx}
\usepackage{textcomp}
\usepackage{subfig}
\usepackage{cleveref}
\usepackage{url}

\makeatletter
\newcommand*\titleheader[1]{\gdef\@titleheader{#1}}
\AtBeginDocument{%
  \let\st@red@title\@title
  \def\@title{%
    \bgroup\normalfont\large\centering\@titleheader\par\egroup
    \vskip1.5em\st@red@title}
}
\makeatother

\usepackage{xcolor}
\def\BibTeX{{\rm B\kern-.05em{\sc i\kern-.025em b}\kern-.08em
    T\kern-.1667em\lower.7ex\hbox{E}\kern-.125emX}}

\titleheader{\tiny This work has been submitted to the 2nd IEEE International Conference on Federated Learning Technologies and Applications (FLTA 2024) for possible publication. Copyright may be transferred without notice, after which this version may no longer be accessible. \\ \vspace{-11mm}}

\title{Exploring Federated Learning Dynamics for Black-and-White-Box DNN Traitor Tracing
\thanks{Work partially funded by the EU (Next Generation) through the "Plan de
Recuperación, Transformación y Resiliencia" under projects FELDSPAR: "Federated Learning with Model Ownership Protection and Privacy Armoring" (Grant MCIN/AEI/10.13039/501100011033)  and TRUFFLES: Trusted Framework for Federated Learning Systems, by the Spanish Ministry of Science, Innovation and Universities via a doctoral grant to the first author (FPU22/01929), and by Xunta de Galicia and the European Regional Development Fund, under project ED431C 2021/47.}
}

\author{\IEEEauthorblockN{Elena Rodríguez-Lois, Fernando Pérez-González}
\IEEEauthorblockA{
\textit{Signal Theory and Communications Department}\\
atlanTTic Research Center, University of Vigo, E.E. de Telecomunicación, 36310 Vigo, Spain\\
{erodriguez, fperez}@gts.uvigo.es\\\vspace{-8mm}}
}

\begin{document}
\maketitle

\begin{abstract}
As deep learning applications become more prevalent, the need for extensive training examples raises concerns for sensitive, personal, or proprietary data. To overcome this, Federated Learning (FL) enables collaborative model training across distributed data-owners, but it introduces challenges in safeguarding model ownership and identifying the origin in case of a leak. Building upon prior work, this paper explores the adaptation of black-and-white traitor tracing watermarking to FL classifiers, addressing the threat of collusion attacks from different data-owners. This study reveals that leak-resistant white-box fingerprints can be directly implemented without a significant impact from FL dynamics, while the black-box fingerprints are drastically affected, losing their traitor tracing capabilities. To mitigate this effect, we propose increasing the number of black-box salient neurons through dropout regularization. Though there are still some open problems to be explored, such as analyzing non-i.i.d. datasets and over-parameterized models, results show that collusion-resistant traitor tracing, identifying all data-owners involved in a suspected leak, is feasible in an FL framework, even in early stages of training.
\end{abstract}


\section{Introduction and Previous Works}

\begin{figure*}[t]
    \centering
    \begin{minipage}[b]{0.3\textwidth}
        \centering
        \subfloat[]{\includegraphics[width=\textwidth]{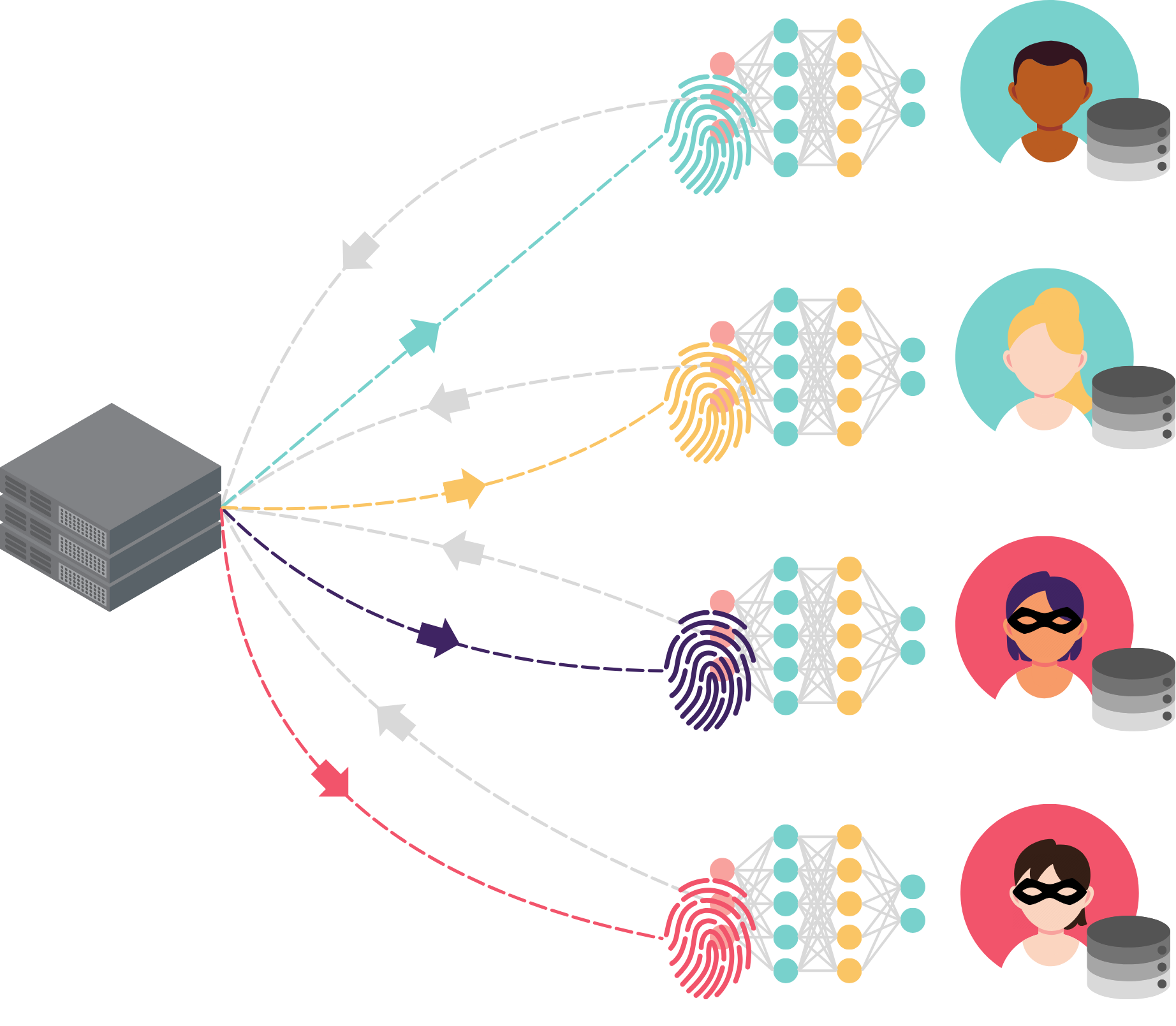}\label{fig:FL_diagram}}
    \end{minipage}
    \begin{minipage}[b]{0.2\textwidth}
        \centering
        \subfloat[]{\includegraphics[width=\textwidth]{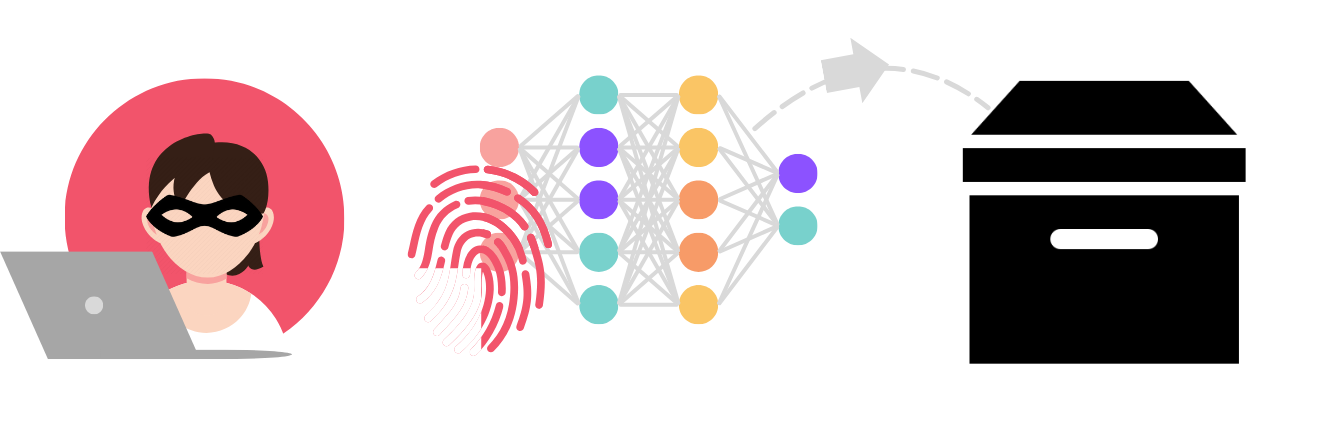}\label{fig:post_training_attack_diagram}}
        \\ \vspace{25pt}
        \subfloat[]{\includegraphics[width=\textwidth]{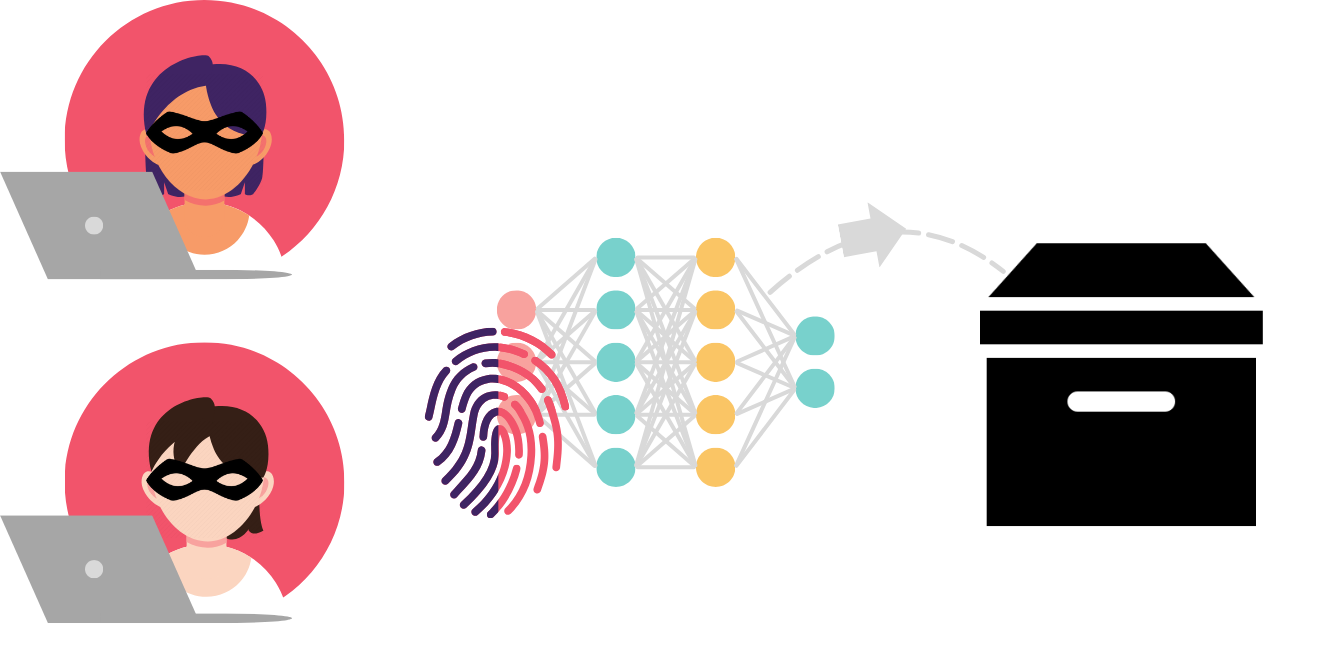}\label{fig:collusion_attack_diagram}}
    \end{minipage}
    \begin{minipage}[b]{0.27\textwidth}
        \centering
        \subfloat[]{\includegraphics[width=\textwidth]{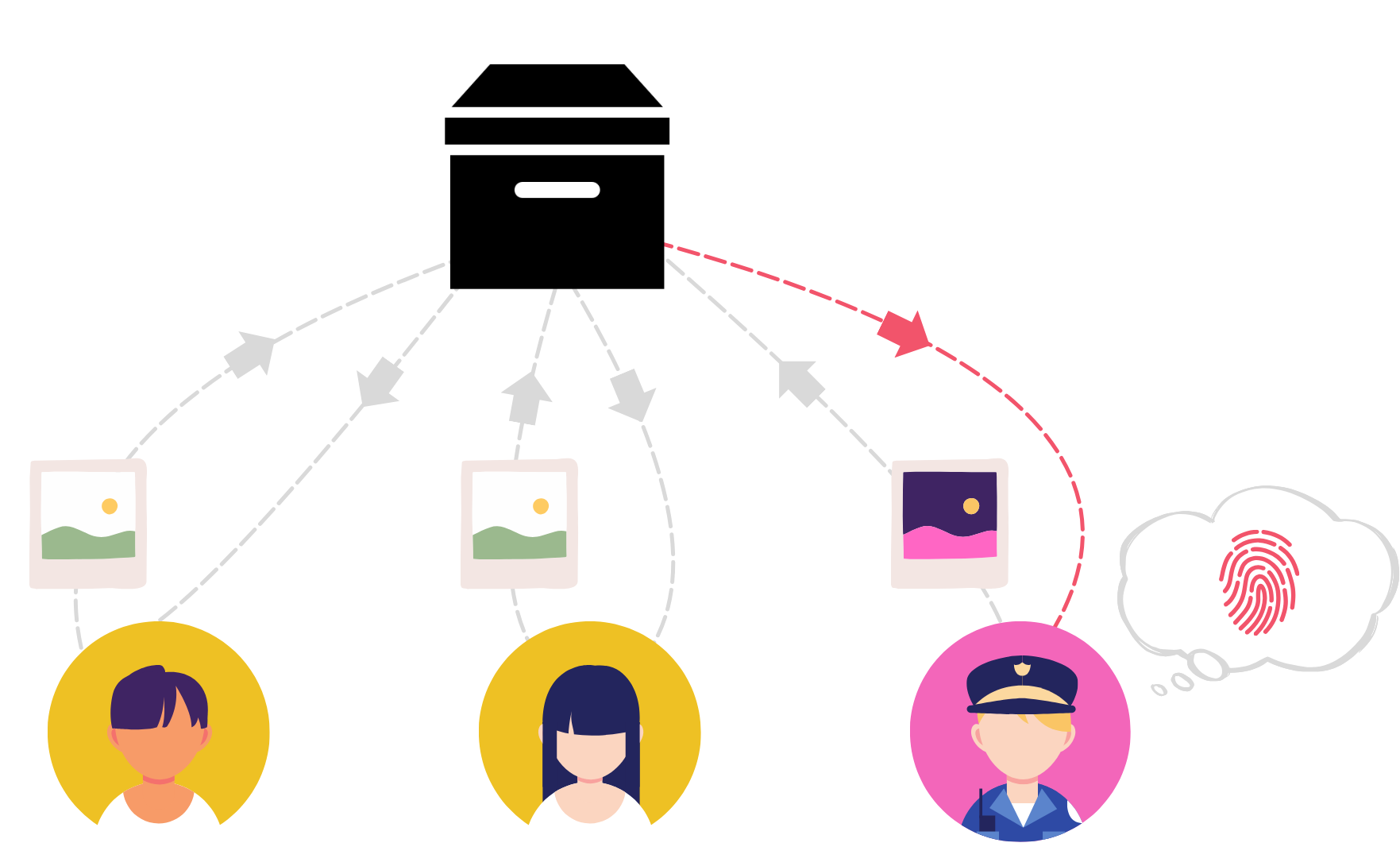}\label{fig:blackbox_verification_diagram}}
        \\ \vspace{10pt}
    \end{minipage}
    \hfill
    \begin{minipage}[b]{0.17\textwidth}
        \centering
        \subfloat[]{\includegraphics[width=\textwidth]{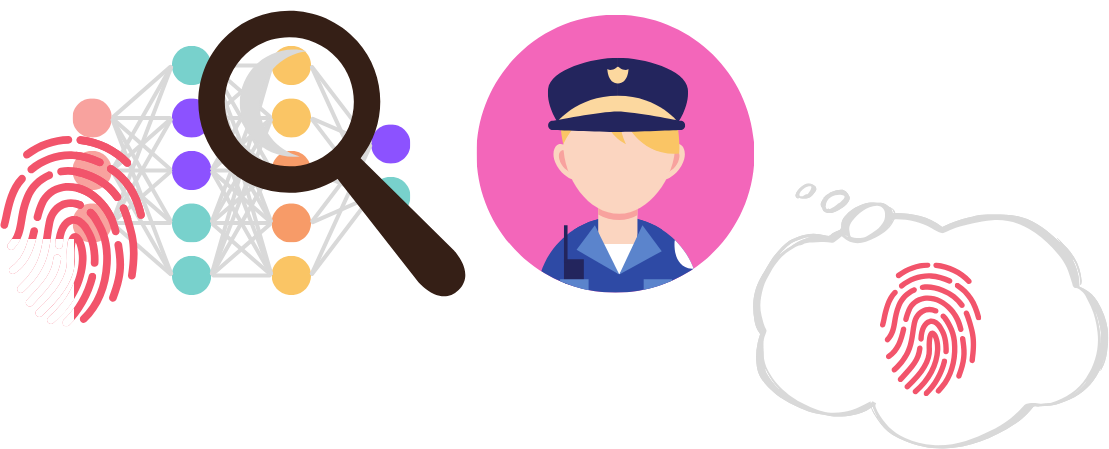}\label{fig:whitebox_verification_1_diagram}}
        \\ \vspace{25pt}
        \subfloat[]{\includegraphics[width=\textwidth]{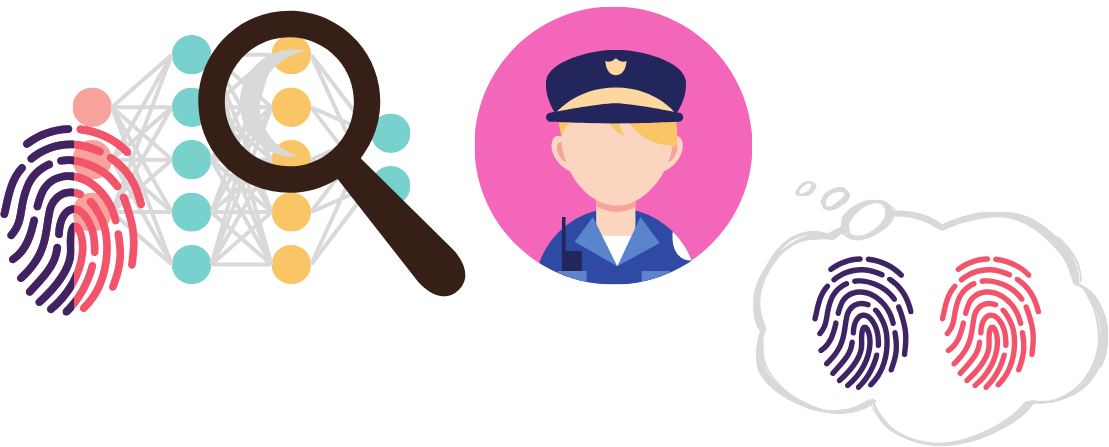}\label{fig:whitebox_verification_2_diagram}}
    \end{minipage}
    \caption{Black-and-white traitor tracing framework for an FL model: training with different fingerprinted models \protect\subref{fig:FL_diagram}, individual \protect\subref{fig:post_training_attack_diagram} and collusion \protect\subref{fig:collusion_attack_diagram} attacks to the watermark, black-box verification through API \protect\subref{fig:blackbox_verification_diagram}, and white-box verification for one \protect\subref{fig:whitebox_verification_1_diagram} or several \protect\subref{fig:whitebox_verification_2_diagram} guilty participants.}
    \label{fig:diagram}
\end{figure*}

With the growing use of deep learning applications, access to large amounts of training examples has become a key factor to achieve highly accurate solutions. This can be difficult to accomplish for some applications, where data may contain sensitive or personal information, and cannot be freely disseminated or shared. In order to manage this constraint, Federated Learning (FL) allows multiple data owners to collaboratively train a machine learning model, without the need to share their private data. Instead, they share only their updates to the model with a central aggregator. This setting, in turn, provides all data owners and the aggregator itself with copies of this shared model, that is susceptible to leakage, espionage or unlawful use. While watermarking has already shown its suitability to deter illegal dissemination of Deep Neural Networks (DNNs) \cite{Li21}, new challenges and scenarios arise that require special consideration in an FL framework. 

For instance, the different participants and potential threats of each application determine which actor in the FL framework carries out the watermarking of the model. In a more direct adaptation of the existing DNN watermarking methods, any or all data-owners can watermark the models locally during their training step, typically allowing them to claim their individual contribution to the overall shared model \cite{Bowen_Li22,Liu21,Yang22,Liang23,WYang23,WYang23-2}. In this case, the data-owner may need to ensure that their watermark survives aggregation, potentially including updates with other watermarks, even after training rounds where they may not be called to participate. Assuming the central aggregator is trusted and has access to the shared model's parameters, it is also possible that they choose to embed a watermark as an additional step \cite{Tekgul21,Fang-Qi_Li21,Shao23,Chen23}, ensuring no data-owner is able to pass the shared model as their sole creation. 

Although there can be many different ways to embed a watermark into a DNN, one can generally distinguish two different approaches, according to the requirements needed to verify its presence. If the information is embedded directly into the internal numeric values of the network, such as the weights or feature maps \cite{Liang23,WYang23,WYang23-2}, verification requires full (white-box) access to the model, which may be unattainable in many cases. Mirroring backdooring attacks, one can also exploit the input-output behavior of the model, and embed the watermark as the output to certain trigger examples \cite{Tekgul21,Liu21,Yang22,Chen23}, making it accessible through model queries (black-box). As both black and white box watermarks present advantages and disadvantages, some works choose to combine them to achieve a stronger scheme \cite{Bowen_Li22,Fang-Qi_Li21,Shao23}. A third alternative is also possible for generative networks where all outputs are watermarked, so no access to the model itself is necessary (box-less) \cite{Fei23}, but this approach is yet to be explored in FL. 

As different watermarks allow the distinction of the different participants involved, some FL works also incorporate a fingerprinting capability, to trace leaks back to their origin. The authors in \cite{Liang23} propose a data-owner-side white-box fingerprinting algorithm, so that it can be implemented alongside secure aggregation for applications where the central aggregator cannot be fully trusted. Given that the watermark embedded for this method contains the participants' fingerprints in each training round, it requires access to the full leaked model weights of several rounds in order to narrow down the accusation to a single data-owner, which may be difficult to achieve, as the culprit may choose a good enough snapshot of the shared model and not update it further. Alternatively, the authors in \cite{Shao23} assume the aggregator is able to produce different watermarked copies of the shared model, and propose a single black-box watermark to prove rightful ownership, alongside white-box fingerprints identifying data-owners individually, once access to the model has been granted. Finally, the work in \cite{Fang-Qi_Li21} suggests the use of distinct black-box fingerprints, comprised of different trigger-label sets memorized on each model copy, that allow direct identification of the leak source from the input-output behavior of the model. However, with a high number of data-owners, this method would require a substantial amount of queries to test triggers from each data-owner's set, which could become time-consuming and potentially expensive. 

Unfortunately, none of these FL works analyze the potential impact of a collusion attack, where multiple participants collaborate to produce a single object, thus complicating the identification of individual watermarks. This attack has been extensively researched in traitor tracing literature, motivated by previous threats such as the unlawful distribution of digital content \cite{Nin13}, and it is very easy to imagine how it would fit into an FL training: a malicious actor could simply pose as two or more data-owners in order to receive different fingerprinted models. While the potential of collusion attacks and collusion-resistant schemes in DNNs is still an emerging field of study, some works have already surfaced attempting to provide a solution to this challenge for independent model copies. So far, these assume a collusion can be expressed as an average of the model weights, and borrow directly from traditional watermarking theory \cite{Chaabane13,Nin13} to overcome this attack. For white-box fingerprints, where the effects of the averaging is straightforward, previous methods have achieved anti-collusion properties through Balanced Incomplete Block Design (BIBD) codes \cite{Chen19} \cite{Xu20}, and also embedding non-binary orthogonal codes, forming a basis of user vectors \cite{RodriguezLois23}. For black-box, the effect of the averaging on the learned triggers is more complex, so our previous work in \cite{RodriguezLois23} exploits the strategy-agnostic Tardos codes to achieve identification of at least one colluder, using a single set of shared triggers with different user labels, minimizing the necessary queries.  

Considering the above, the goal of this paper is to adapt the work in \cite{RodriguezLois23} to an FL framework, and to understand the new dynamics, changes and limitations that emerge as a result.

In what follows, \Cref{sec:problem} presents the problem of FL watermarking and discusses the two main challenges w.r.t. the watermarking of independent models, \Cref{sec:proposed} introduces the black-and-white traitor tracing scheme from \cite{RodriguezLois23}, \Cref{sec:experimental} presents the implementation details and the discussion of experimental results, and finally \Cref{sec:conclusion} closes this paper with the main conclusions and open challenges. 

\section{Watermarking FL models} \label{sec:problem}

In traditional deep learning schemes, the owner is able to watermark their model to protect its IP and prevent theft, trusting they are the only one that could have accessed a non-watermarked version of the weights, and that any malicious actor will need to attack the watermarked copy in order to remove the watermark, compromising the performance of the model on the main task. Although they may have white-box access to the stolen model, the attacker would only have the final snapshot of the model weights, and limited knowledge about the main task (such as a smaller version of the training dataset) and the watermarking scheme (such as the embedding function). An FL framework, on the other hand, allows for many participants, including all data owners and potentially the aggregator, to access the training model at many stages of the process, any of which could choose to act maliciously against the system. 

Although some works have been presented that try to prevent or mitigate Byzantine attacks also through the watermarking of the shared model \cite{Zheng22}, our work focuses on model theft and unlawful dissemination, assuming every party, even the attacker, has it in their best interest that the performance of the model is not compromised. Still, different threat models could be defined depending on which party or parties are deemed to be untrustworthy. 

This paper will assume an FL framework with a trusted aggregator, as described in \Cref{fig:diagram}. For each data-owner's model copy, the aggregator will embed distinct black and white-box fingerprints, to deter them from exploiting the shared model outside the FL agreement (\ref{fig:FL_diagram}). At any point in the training process, any data-owner may attempt to remove this fingerprints, either acting alone (e.g., fine-tuning, pruning) or in collusion with other data-owners (e.g., model averaging) (\ref{fig:post_training_attack_diagram}, \ref{fig:collusion_attack_diagram}), while trying to preserve the functionality of the FL model as much as possible. Assuming a leak has taken place, the identification of the black-box fingerprint could be performed against the operating leaked model through queries to an API (\ref{fig:blackbox_verification_diagram}), resulting in the accusation of one of the participants involved and allowing further legal action. Finally, after access to the full model is granted, the white-box fingerprint can be used to gather further proof and identify other data-owners involved (\ref{fig:whitebox_verification_1_diagram}, \ref{fig:whitebox_verification_2_diagram}).

To achieve the usual considerations of \textit{Robustness}, \textit{Security}, \textit{Fidelity}, \textit{Capacity}, \textit{Integrity}, \textit{Generality} and \textit{Efficiency} (defined in \cite{Li21}), this configuration raises two main challenges that did not apply to our previous work in \cite{RodriguezLois23}: 

\paragraph{Fingerprint leak through model updates} In traditional FL frameworks, all data-owners train their local dataset on the same shared model, and any update to this model contains only information about the local datasets, promoting a better performance on the main task for all participants. In this case however, as data-owners use a watermarked model copy to perform their local training, it is likely that the model updates they share are biased, and contain also information of the watermark itself, which would entail that model copies are not only affected by their own fingerprint, but also by those of others, potentially leading to fingerprint leaks. This effect was briefly mentioned in \cite{Shao23}, where the authors chose to stop the insertion of the white-box fingerprint once verification could be achieved, in hopes of reducing its impact on model updates. Although certain white-box schemes could help mitigate this effect (see \Cref{subsec:white}), it is still an open challenge for back-box watermarks, where the embedding space most resembles the main task's. 

\paragraph{Dynamics of the main task compared to the traitor-tracing capabilities} Given that data-owners have access to many intermediate snapshots, the traitor tracing capabilities must already be present once the model reaches a desired performance on the main task, to prevent a malicious actor from settling for a less than optimal, untraceable model. Unfortunately, the works in \cite{Shao23,Fang-Qi_Li21} do not explore this issue, and consider only the watermarking accuracy after the training is done. Generally speaking, as the definition of desired performance may change from one application to another, one could set a conservative threshold assuming any improvement of the FL system must be protected. This means that as soon as the collaborative model begins to outperform what a single data-owner could achieve independently, the framework should ensure that traitor tracing mechanisms are already in place to protect against potential leaks or unlawful use.

\section{Watermarking Scheme} \label{sec:proposed}
Drawing from our previous work in \cite{RodriguezLois23}, the current approach will leverage the practicality of the black-box traitor tracing and the high capacity of white-box fingerprints to allow identification of all data-owners involved in the leak of a DNN classifier, even if the suspected model is initially only accessible through an API. For any collusion of $c$ data-owners, defined by the set of their indexes $\mathcal{C}^c$, the black-box scheme should achieve the identification of at least one guilty data-owner (\textit{catch-one} goal), allowing further action and the request of the full model weights for inspection. The white-box scheme should then serve as greater evidence, identifying the remaining data-owners involved in the leak (\textit{catch-all} goal).

\subsection{Black-Box Fingerprinting with $q$-ary Tardos Codes} \label{subsec:black}
The strategy-agnostic, collusion-resistant capabilities of Tardos Codes \cite{Skoric12} can be exploited to generate different data-owner labels to a set of $m$ shared triggers $\mathcal{T}^m$ \cite{RodriguezLois23}. For each data-owner $j$, this would be a vector $\textbf{x}_j$, where each shared trigger $i$ is assigned a $q$-ary label $x_{ji}$. This allows the utilization of the full dimensionality of the output, by setting $q$ to match the number of classes in the classifier. These labels $x_{ji}$ are randomly sampled, and will be assigned a certain output class $\alpha$ with probability $P[x_{ji}=\alpha] = p_\alpha^{(i)}$ according to a secret $q$-component bias vector $\textbf{p}^{(i)}$, independently drawn from a from a symmetric Dirichlet distribution, with concentration parameter $\kappa>0$ and cutoff parameter $\tau$, ensuring $p_\alpha \in [\tau, 1-(q-1)\tau]$ \cite{Skoric12}. These traitor tracing codes rely on the Marking Assumption (MA), which in this scenario would require that the merged model is only able to output, for a certain trigger, one of the labels assigned to the colluders. While in \cite{RodriguezLois23} it was shown that the MA would not always hold, the scheme proposed therein was still successful by exploiting the information in the surviving triggers.

As done in \cite{RodriguezLois23}, the accusation is an iterative process, where triggers are sequentially fed to the network. At any given point, the suspected model has answered $t$ queries, with each new trigger $i$ updating the cumulative score $S_j^t$ for each data-owner as
\begin{equation} \label{eq:cumscore}
    S_j^t=\sum_{i=1}^t S_j^{(i)},
\end{equation}
where $S_j^{(i)}$ is calculated according to the output to the trigger $i$ and the data-owner labels $x_{ji}$ 
\begin{equation} \label{eq:segmentscore}
    S_j^{(i)}= 
        \begin{cases}
        U_1(p^{(i)}_{y_i}) & \text{if } x_{ji} = y_i \\
        U_0(p^{(i)}_{y_i}) & \text{if } x_{ji} \neq y_i
        \end{cases}   ,
\end{equation}
where 
\begin{align} \label{eq:scorefunctions}
    U_1(p)&=\sqrt{(1-p)/p},  & U_0(p)&=-\sqrt{p/(1-p)}.
\end{align}
This process will end after $t^*$ queries, whenever any of the data-owners' score $S_j^{t^*}$ surpasses an accusation threshold $Z_{t^*}$. From the work in \cite{Skoric12}, a conservative $Z_t$ can be set ensuring a false positive rate below a certain $\epsilon$:
\begin{equation} \label{eq:pfpbound}
    \epsilon \leq \text{exp}\left(-\frac{Z_t^2}{2t} \cdot \frac{1}{1 + Z_t / (3t\sqrt{\tau})}\right),
\end{equation}
meaning that after $t$ queries, threshold $Z_t$ will be a positive number that can be computed as 
\begin{equation} \label{eq:quad}
    Z_t=\ln{\epsilon}\left(-\frac{1}{3\cdot \sqrt{\tau}}\pm\sqrt{\frac{1}{9\cdot \tau}-\frac{2t}{\ln{\epsilon}}}\right).
\end{equation}

Our work in \cite{RodriguezLois23} also used Sequential Probability Ratio Test (SPRT) \cite{SPRT} thresholds that allow the accusation process to stop early also for a suspected model that was not actually derived from any of the protected copies, but, in contrast to the simplicity of $Z_t$, these require the experimental modeling of scores $S_j^{(i)}$. Although it is possible to implement the SPRT in the FL framework, it could become computationally expensive as models, and thus potentially the distribution of $S_j^{(i)}$, are constantly updated.

\subsection{White-Box Fingerprinting with Orthogonal Codes} \label{subsec:white}
As done in \cite{RodriguezLois23}, a white-box fingerprint can be embedded alongside the black-box watermark, exploiting the over-parametrization of the model to encode the identifying information into the numerical values of the model weights. Considering a vector $\textbf{w}$ containing the $l$ flattened weights of a certain layer, fingerprinting can be achieved by adding a regularization term $E_R$ to the total loss $E$, such as
\begin{equation}
    E(\textbf{w}) = E_0(\textbf{w}) + \lambda E_R(\textbf{w}),
\end{equation}
where $E_0$ is the classification loss, and $\lambda$ is a constant controlling the strength of the watermark. The different data-owners' fingerprints are generated as a $p$-dimensional orthonormal basis $\mathcal{S} = \{\textbf{s}_1, \textbf{s}_2, ..., \textbf{s}_p\}$, with $\textbf{s}_j$ assigned to data-owner $j$, and a secret matrix $\textbf{D}^{l \times p}$ is randomly sampled from a normal distribution, projecting the $l$ elements of vector $w$ into the data-owner vector $\textbf{s}_j$ in $\mathcal{S}$. This method ensures that the embedding for a certain data-owner also cancels any leak from other fingerprints, as opposed to the one used in \cite{Shao23}, where unique projection matrices were used. The projection over the basis and the regularization loss can then be computed as
\begin{align} \label{eq:proj}
    r_j &= \frac{\textbf{w}^\intercal \cdot \textbf{D} \cdot \textbf{s}_j}{\|\textbf{w}^\intercal \cdot \textbf{D}\|}& &\text{and}  & E_R(\textbf{w}) &= \exp \left(-r_j\right)
\end{align}
respectively, promoting that the projected vector aligns with the desired $\textbf{s}_j$. For a collusion $\mathcal{C}^c$, the expectation of the projection for a guilty and an innocent data-owner would be
\begin{align} \label{eq:eproj}
    \mathbb{E} \{r_{j|j\in \mathcal{C}^c}\} &= 1/\sqrt{c}& &\text{and}  & \mathbb{E} \{r_{j|j\notin \mathcal{C}^c}\} &= 0.
\end{align}

\section{Implementation and Experimental Results} \label{sec:experimental}

\subsection{DNN Architecture and Main Task}

In order to better isolate the impact of the FL dynamics on the traitor tracing capabilities of the scheme, the experiments described in this work follow the same configuration as those presented in \cite{RodriguezLois23}. Specifically, we will consider an image classification task on the MNIST dataset \cite{MNIST} with a small DNN architecture, which is described in \Cref{tab:architecture}, as a toy example. For all purposes, the output class $y_i$ will be considered as the highest-value output neuron, regardless of the soft value of the vector.

\begin{table}
    \centering
    \begin{tabular}{|c|c|c|}
        \hline
        Layer & Size & Activation \\ \hline
        Convolutional layer 1 & 16 kernels 3$\times$3 & ReLu \\ \hline
        Convolutional layer 2 & 64 kernels 3$\times$3 & ReLu \\ \hline
        Convolutional layer 3 & 128 kernels 3$\times$3 & ReLu \\ \hline
        Fully connected layer & 10 neurons & Softmax \\ \hline
    \end{tabular}
    \caption{Small DNN architecture.}
    \label{tab:architecture}
\end{table}

\subsection{Watermarking Configuration}
As with the main task, the configuration of the black and white-box watermark borrows from the findings in \cite{RodriguezLois23} to aid this analysis:

\paragraph{Black-box fingerprints} The shared trigger set $\mathcal{T}^m$ was generated by independently sampling from a uniform distribution in the range [0,1],\footnote{Although the work in \cite{RodriguezLois23} also evaluates the use of benign triggers, the embedding space in such case would be even closer to the main task, which could complicate the analysis of the potential fingerprint leaks.} with size $m=1000$. The maximum number of colluders considered $c_0$, with $\mathcal{C}^c$, $c<c_0$, is 6, the cutoff parameter $\tau$ is 0.038, and the concentration parameter $\kappa$ is set to 100, which favours majority-voting strategies, and was shown to be the most effective in \cite{RodriguezLois23}. The maximum false positive rate $\epsilon$ considered for the threshold $Z_t$ is $10^{-6}$. 

\paragraph{White-box fingerprints} Following the configuration in \cite{RodriguezLois23}, the third convolutional layer with 73,728 parameters is chosen to embed the white-box watermark, with a data-owner basis of $p=1000$ dimensions, and the strength of the regularization term $\lambda$ set to 1.

\subsection{Training Strategies} \label{subsec:strategies}
For all experiments, 25\% of the MNIST dataset (17,500 images) is reserved for the evaluation of the main task and as additional data for the data-owners' attacks. The remaining subset for training (52,500 images) is equally distributed among 100 data-owners. The different training strategies used in the experiments can be described as follows:

\paragraph{No WM} For each training round, the aggregator will randomly choose 10 data-owners, who will provide the gradients for a mini-batch of 16 local examples with categorical cross-entropy loss. All 10 gradients are averaged, and the shared model is updated using stochastic gradient descent with a learning rate of 0.001. This is repeated through 5 epochs (1640 iterations total).
\paragraph{Vanilla WM} The aggregator generates 100 fingerprinted model copies, one for each data-owner, and updates them individually on each round. After the main task optimization as described in the No WM strategy, the aggregator will perform an additional black-and-white watermarking step, forcing each copy from data-owner $j$ to both memorize the shared trigger set $\mathcal{T}^m$ with the desired labels $\textbf{x}_j$, and align their weights with their corresponding data-owner vector $\textbf{s}_j$ through the regularization term. As with the main task, this step uses a mini-batch size of 16, categorical cross-entropy loss, and a learning rate of 0.001.
\paragraph{Dropout WM} As noted in previous works \cite{Chen22}, the memorization of black-box triggers may rely on a very small subset of salient neurons, which is a threat to the robustness of the scheme. Furthermore, the leaks of the black-box fingerprints may be promoting an even smaller subset (see \Cref{subsec:results}), as any neuron relevant to the main task is more likely to be disturbed in the shared updates, thus delegating the memorization of the triggers to weak neurons. To mitigate this effect, and building on the Vanilla WM, this strategy utilizes dropout layers with probability 0.2 before every layer on the network. 
\paragraph{Dropout \& Limited WM} To address the \textit{Efficiency} concerns introduced by the additional watermarking step, which unfortunately imposes a non-negligible computational overhead, one could implement a flexible strategy aiming to skip this step whenever it is not necessary. Initially, it is crucial for the fingerprints to become detectable early in the training process, so the watermarking step is performed on every iteration during the first two epochs. This ensures that traitor tracing capabilities are established before the model becomes susceptible to theft, as discussed in \Cref{sec:problem}. As the fingerprints become more robust in the model copies, the frequency of watermarking is reduced. It is performed every 10 iterations for the next two epochs, and only every 100 iterations in the final epoch. This limitation on the watermarking step reduces the computational overhead in the later stages of training while ensuring early detection of fingerprints.
\paragraph{Independent Models} As the main task learning and watermarking step are separate in the FL framework, a direct comparison to the models trained in \cite{RodriguezLois23} may overlook other factors affecting the traitor tracing capabilities of the scheme, so independent models were also trained, assuming access to the full MNIST dataset, emulating the training rounds in Vanilla WM. So for each training round, the model updates its weights according to a main task batch with 160 examples, and later performs the watermarking step with the black-and-white fingerprints.
\paragraph{Independent Data-Owner Models} Models were also trained independently with the individual data-owners subsets, without any watermark and batch size 16, to find a conservative threshold on the main task accuracy after which model copies should be protected, as mentioned in \Cref{sec:problem}.
\paragraph{Vanilla Dif WM and Dropout Dif WM} The Vanilla WM and Dropout WM strategies were also tested with different sets of triggers for each data-owner $\mathcal{T}_j$, to ensure that the impact of FL on the traitor tracing capabilities of the fingerprints is not unique to shared triggers. 

\subsection{Data-Owner Attacks on the Fingerprints}
The collusion attack, where $c$ data-owners $\mathcal{C}^c$ merge their model copies to attack their individual fingerprints, is the main focus of this work. Although other implementations may be possible, this work will consider the collusion as the averaging of the parameters, as was done in \cite{RodriguezLois23}. Additionally, after the merged copy has been generated, data-owners could further attack this model by fine-tuning or pruning its weights. In these experiments, a fine-tuning attack will train on the validation examples from MNIST for 5 epochs, and a pruning attack will use the same images to set 80\% of the least relevant neurons to 0. Other possible attacks should be considered in future works.

\subsection{Results} \label{subsec:results}

\begin{figure}[t]
    \centering
    \subfloat[Main task]{\includegraphics[width=0.45\columnwidth]{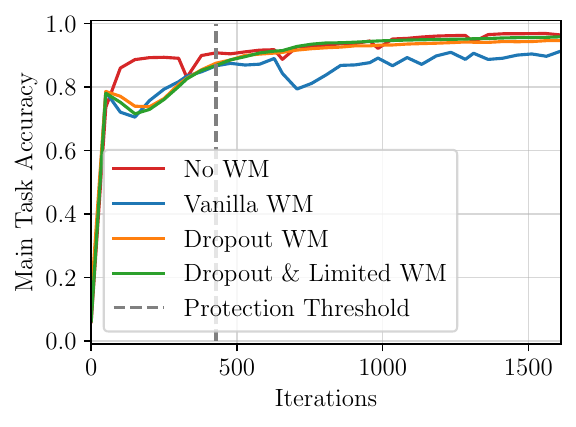}\label{fig:training_evolution}}
    \subfloat[White-box WM]{\includegraphics[width=0.45\columnwidth]{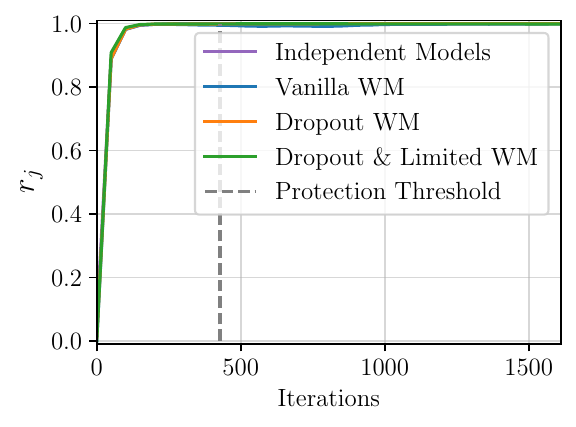}\label{fig:projection_evolution}}\\
    \subfloat[Black-box WM]{\includegraphics[width=0.45\columnwidth]{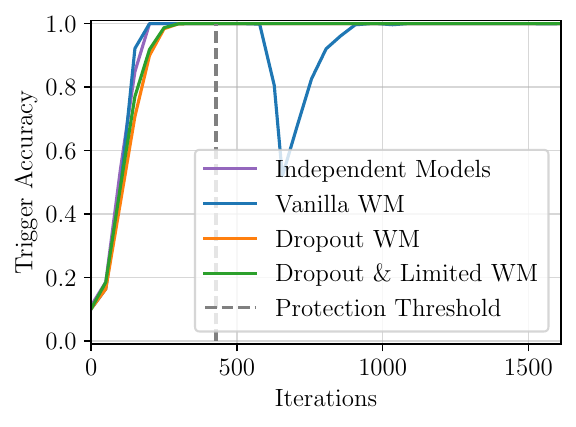}\label{fig:trigger_evolution}}
    \subfloat[Black-box MAV for $\mathcal{C}^2$]{\includegraphics[width=0.45\columnwidth]{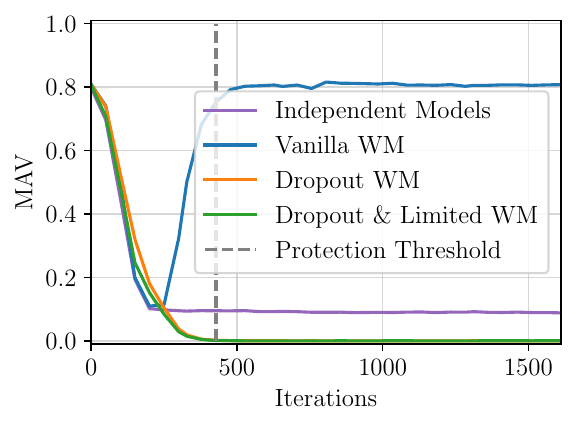}\label{fig:mav_evolution}}
    \caption{Training evolution of FL models.}
    \label{fig:evolution}
\end{figure}

The evolution of the main task and watermarking for the different training strategies can be seen in \Cref{fig:evolution}, averaging the metrics of 10 random models out of the 100 data-owners, and indicating the protection threshold after the watermarked  models (Vanilla WM, Dropout WM, Dropout \& Limited WM) surpass the main task accuracy achievable by data-owners on their own, an average of 85.91\%. As the white-box fingerprints are leak-resistant by design, no impact can be seen on \Cref{fig:projection_evolution}, and the projection is correctly embedded before the protection threshold in any case. For the black-box fingerprint, even though all strategies seem to accurately classify the shared trigger set, for the most part, after the threshold in \Cref{fig:trigger_evolution}, this does not necessarily ensure the traitor tracing capabilities of the scheme. The work in \cite{RodriguezLois23} also measures the violations to the MA (MAV), defined by the collusion outputting a class that is different to the colluders' labels, as 
\begin{equation}
    \text{MAV} = \frac{1}{m} \sum_{i=1}^{m} [y_i \notin \mathcal{X}_{\mathcal{C}i}],
\end{equation}
where $\mathcal{X}_{\mathcal{C}i}$ is a set of all $x_{ji}$ for data-owners $j \in \mathcal{C}$, and $[\cdot]$ represents the Iverson bracket. Now, according to the experimental MAV on 10 random collusions $\mathcal{C}^2$ in \Cref{fig:mav_evolution}, one can see that after reaching 100\% accuracy on the triggers, the Vanilla WM ends up overfitting their fingerprints to internal features that do not survive the collusion, and thus, are not compatible with traitor tracing. Additionally, the dropout regularization on the watermarking step seems to have a positive effect on the convergence of the main task in \Cref{fig:training_evolution}, while the Vanilla WM has a noticeable impact on the models' accuracy.

\begin{figure}[t]
    \centering
    \subfloat[Black-box WM]{\includegraphics[width=0.45\columnwidth]{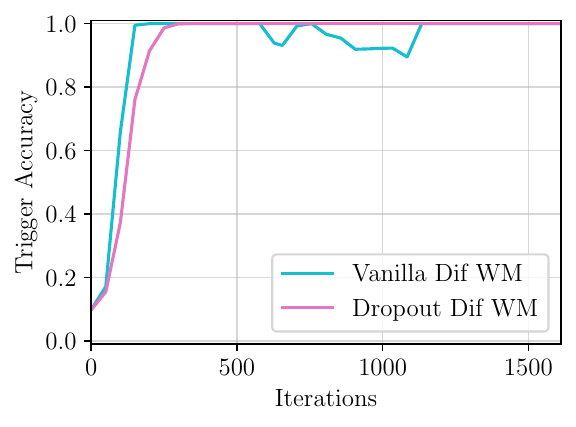}\label{fig:trigger_evolution_FQ}}
    \subfloat[Black-box WM for $\mathcal{C}^2$]{\includegraphics[width=0.45\columnwidth]{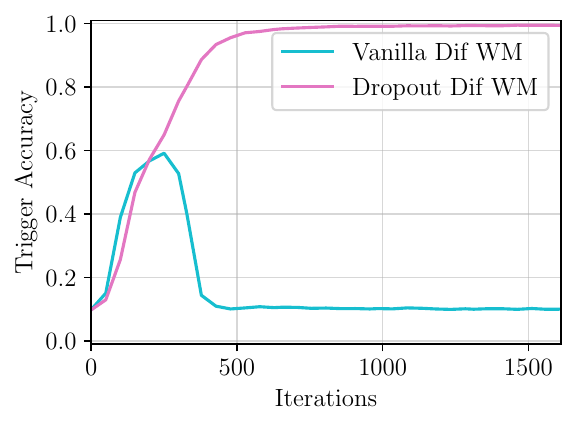}\label{fig:col_evolution_FQ}}
    \caption{Training evolution of FL models with different trigger sets $\mathcal{T}_j$.}
    \label{fig:evolution_FQ}
\end{figure}

Unfortunately, this vulnerability to the collusion attack is not exclusive to our proposed scheme, and one can see a similar effect on \Cref{fig:evolution_FQ} if each data-owner $j$ is assigned unique trigger-label pairs $(\mathcal{T}_j, \textbf{x}_j)$, as done in \cite{Fang-Qi_Li21}. Considering this, it is more efficient to use the same trigger set $\mathcal{T}$ across all data-owners, which significantly reduces the number of necessary queries before an accusation \cite{RodriguezLois23}.

\begin{figure}[t]
    \centering
    \includegraphics[width=\columnwidth]{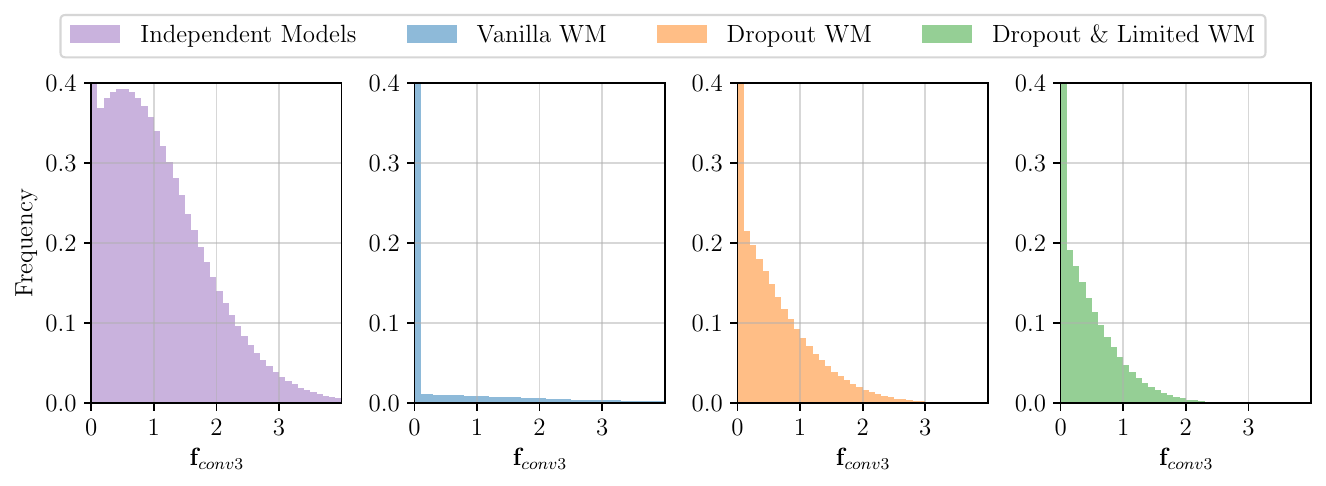}
    \caption{Histogram of features $\textbf{f}_{conv3}$ for the trigger set $\mathcal{T}$.}
    \label{fig:hist_features}
\end{figure}

Exploring the hypothesis of \Cref{subsec:strategies}, where black-box triggers in Vanilla WM could potentially be delegated to a small, weak, subset of neurons, \Cref{fig:hist_features} shows the histogram of the flattened outputs $\textbf{f}_{conv3}$ of the third convolutional layer for different training strategies, after feeding a given data-owner's network with the trigger set $\mathcal{T}$. It is evident that the number of activations greater than 0.1 for the Vanilla WM is extremely low when compared to the Independent Models, and that dropout regularization aids in the increase of salient neurons, making for a more robust fingerprint. 

\begin{figure}[t]
    \centering
    \subfloat[No further attacks]{\includegraphics[width=\columnwidth]{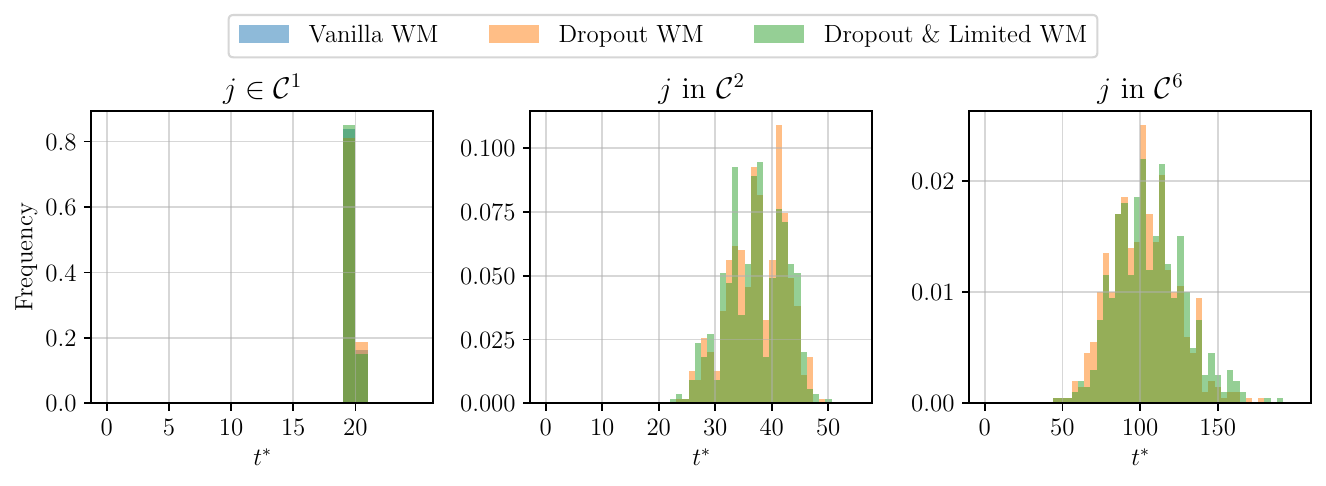}\label{fig:m_needed}}\\
    \subfloat[Fine-tuning]{\includegraphics[width=\columnwidth]{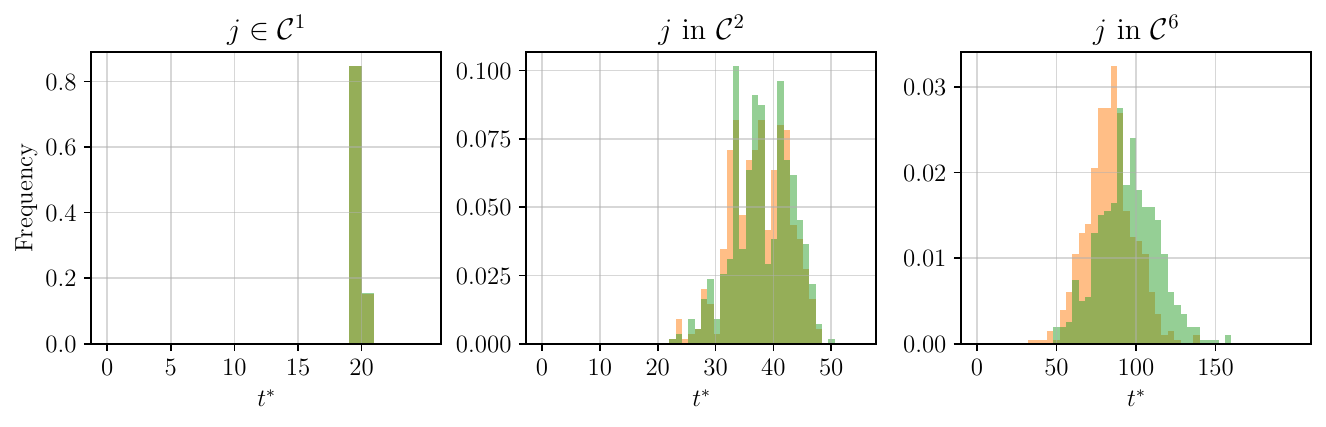}\label{fig:m_needed_fine}}\\
    \subfloat[Pruning]{\includegraphics[width=\columnwidth]{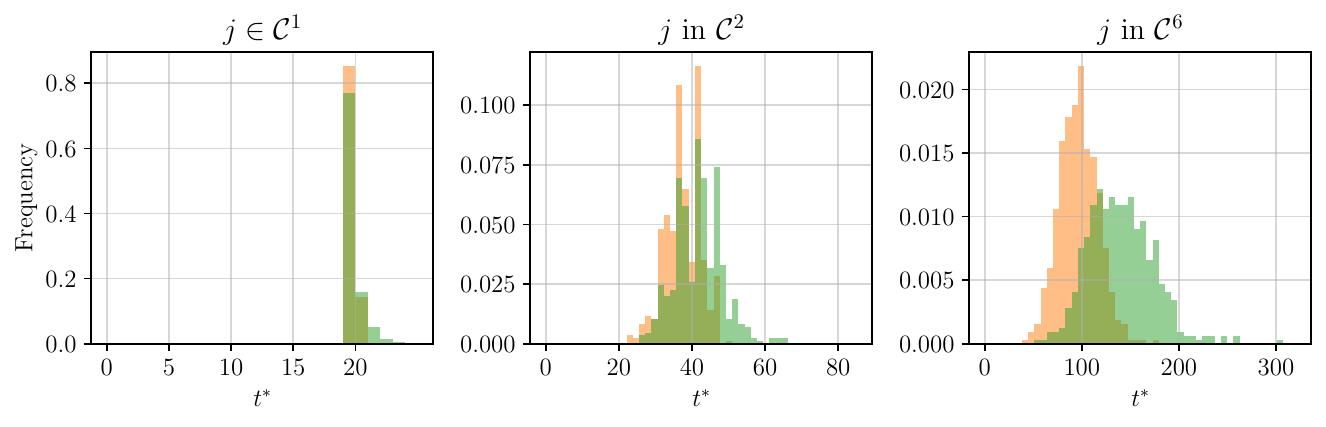}\label{fig:m_needed_prun}}
    \caption{Experimental distribution of $t^*$ according to different DNN attacks after collusion.}
    \label{fig:m_needed_all}
\end{figure}

In terms of the black-box accusation process, \Cref{fig:m_needed_all} shows the minimum number of queries $t^*$, out of the available $m$ embedded triggers, required before an accusation can be made on a particular colluder, on 500 collusions randomly chosen from all 100 data-owners. Both Dropout WM and Dropout \& Limited WM are able to catch a colluder using less than half of the available triggers, even when pruning 80\% of the neurons. The repetition of the watermarking step in Dropout WM seems to have made it more robust to attacks, especially pruning, but depending on the application, the computational overhead may shift the balance in favor of Dropout \& Limited WM, as the difference in $t^*$ is not too extreme. For a guilty data-owner $j\in\mathcal{C}^6$ and pruning of the merged model, the average $t^*$ is 97 and 140 respectively. While the Dropout and Dropout \& Limited WM had no false negatives in the generated collusions, the same cannot be said about Vanilla WM, as can be seen in \Cref{tab:fn_vanilla}. As expected, this strategy was unable to detect a single guilty participant in the majority of cases, unless the models were not attacked at all (i.e. only one guilty participant $j\in\mathcal{C}^1$ with no further attacks), showing that the direct implementation of the work in \cite{RodriguezLois23} is not possible for an FL framework, unless one considers the potential effect of the fingerprint leaks on the MAV.

\bgroup
\def\arraystretch{1.25}
\begin{table}
    \centering
    \begin{tabular}{c|c|c|c|}
        \cline{2-4}
        & $j\in\mathcal{C}^1$ & $j\in\mathcal{C}^2$ & $j\in\mathcal{C}^6$ \\ \hline
        \multicolumn{1}{|r|}{No further attacks} & \textbf{0} (\textbf{0\%}) & 497 (99.4\%) & 500 (100\%)\\ \hline
        \multicolumn{1}{|r|}{Fine-tuning} & 421 (84.2\%) & 438 (87.6\%) & 500 (100\%) \\ \hline
        \multicolumn{1}{|r|}{Pruning} & 359 (71.8\%) & 412 (82.4\%) & 500 (100\%) \\ \hline
    \end{tabular}
    \caption{Number of false negatives for Vanilla WM in 500 random collusions.}
    \label{tab:fn_vanilla}
\end{table}
\egroup

\begin{figure}[t]
    \centering
    \subfloat[No further attacks]{\includegraphics[width=\columnwidth]{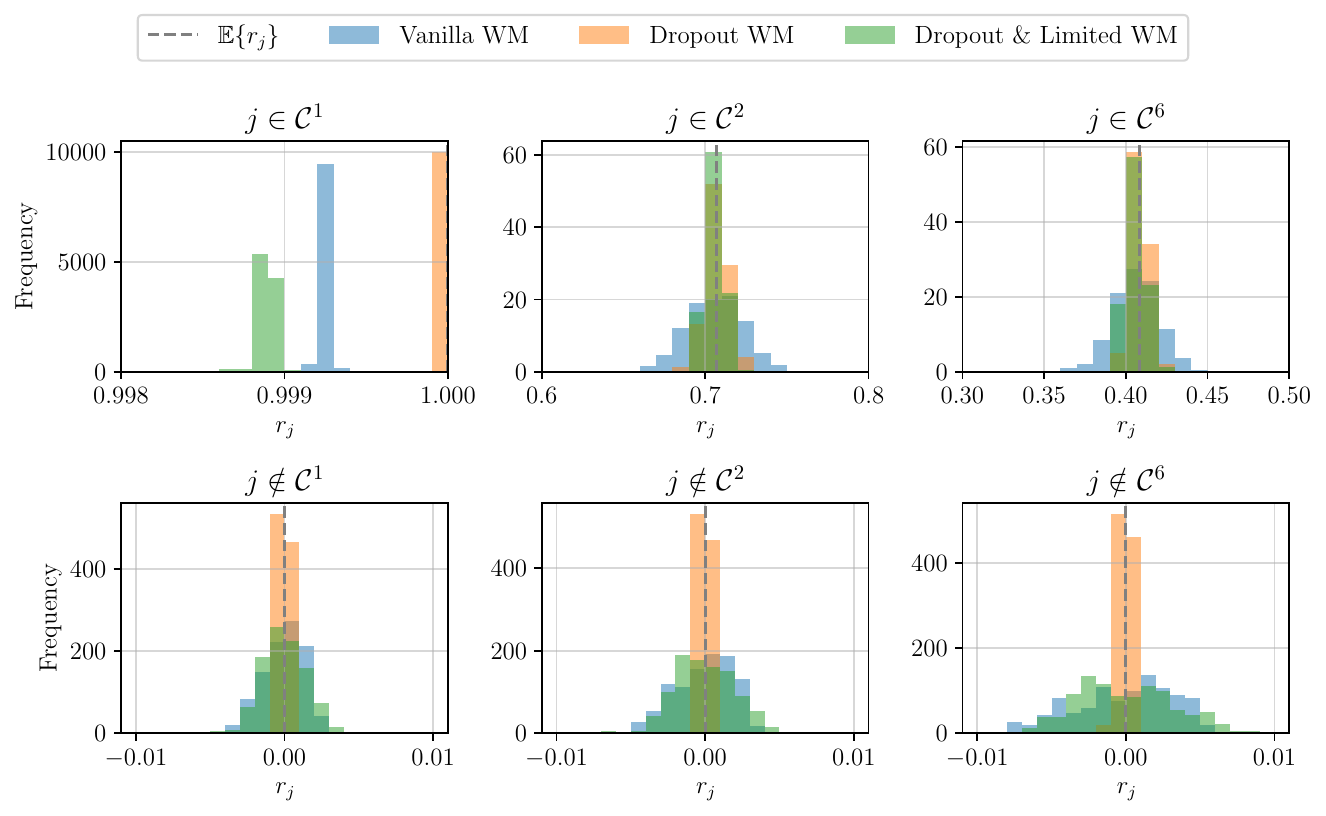}\label{fig:pwb}}\\
    \subfloat[Fine-tuning]{\includegraphics[width=\columnwidth]{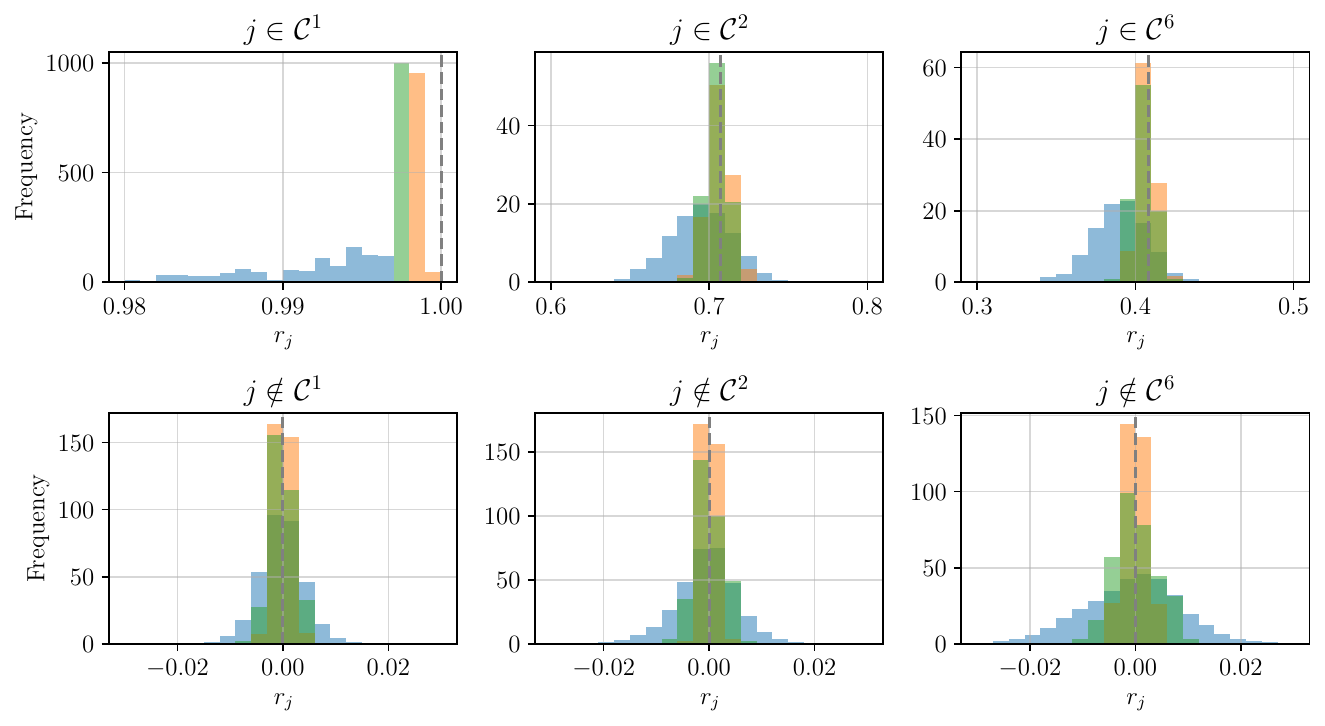}\label{fig:pwb_fine}}\\
    \subfloat[Pruning]{\includegraphics[width=\columnwidth]{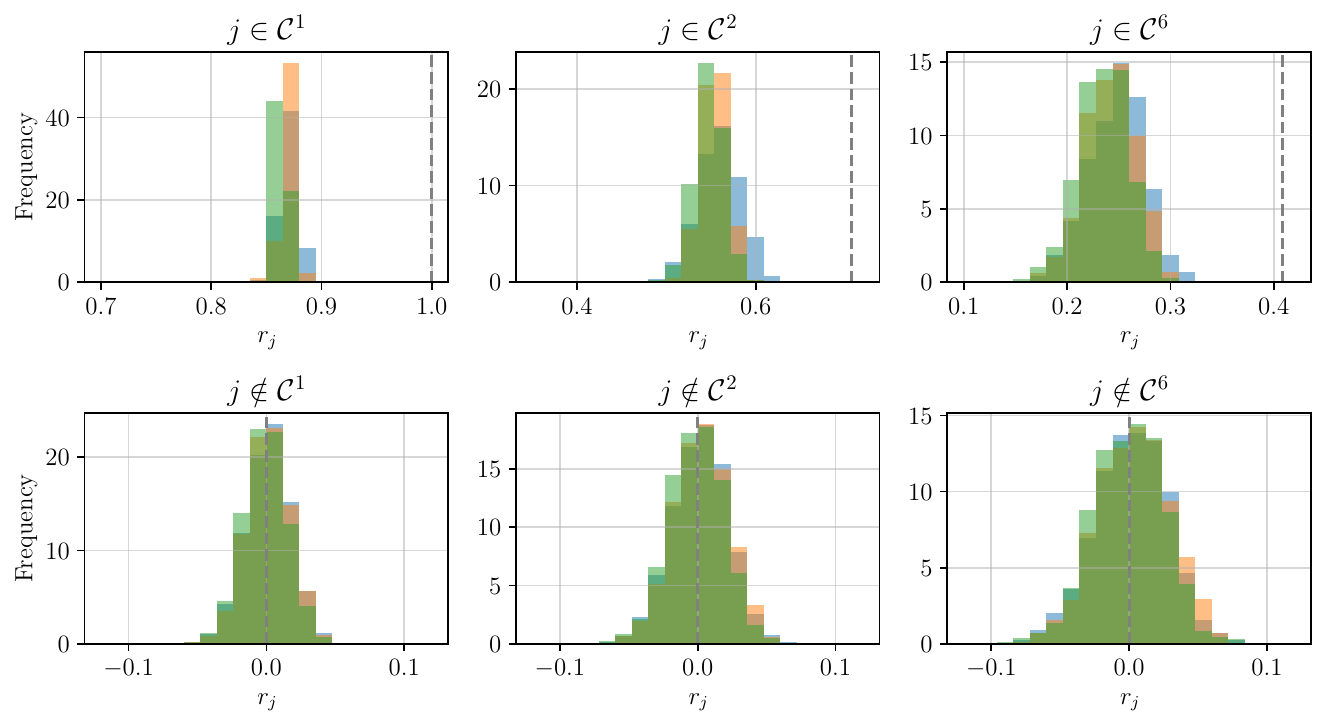}\label{fig:pwb_prun}}
    \caption{Experimental distribution of $r_j$ according to different DNN attacks after collusion.}
    \label{fig:pwb_all}
\end{figure}

For the white-box accusation, the projection $r_j$ of the model weights over the data-owner basis $\mathcal{S}$ can be seen in \Cref{fig:pwb_all}, for 500 random collusions of the 100 data-owners. Although there is a slight benefit to the Dropout WM strategy, as the dropout regulatization promotes more neurons to contribute to the alignment of the weights to the data-owner vector $\textbf{s}_j$, this benefit is not too significant on the practicality of the scheme. Still, in the most challenging scenario considered (with guilty data-owners $j\in\mathcal{C}^6$ and an additional pruning attack), and setting an accusation threshold of 0.11 to avoid false accusations, all strategies would achieve their \textit{catch-all} goal in all the cases considered.

\section{Conclusion} \label{sec:conclusion}
While the threat of data-owner collusion is viable in the context of FL, it had not been studied whether previous collusion-resistant traitor tracing methods are compatible with the training dynamics of an FL framework. A first step toward understanding their impact is presented in the current work, based on our previous black-and-white watermarking scheme in \cite{RodriguezLois23}. Since the original design for the white-box fingerprint is inherently leak-resistant, there were no significant complications with the shared updates when embedding the different model copies. In contrast, implementing the black-box embedding directly on the FL framework would not provide traitor tracing capabilities, most likely because very few neurons are allocated to memorize the fingerprints, making them weak and vulnerable to collusion attacks. With the proposed solution, which promotes an increase in salient neurons through dropout regularization, the FL traitor tracing framework would finally be able to accurately identify all data-owners behind a suspected leak, even at early stages of training, when the main task accuracy might already be appealing for a malicious participant.  

While these results show promise, several challenges and limitations remain to be addressed in future research. Firstly, while reducing the computational overhead by restricting the watermarking step to specific training rounds in later stages is feasible, it comes at a slight cost, weakening the fingerprints' robustness to attacks. Ensuring a timely embedding of watermarks also necessitates prior knowledge of the main task dynamics, which may not be feasible in certain applications. Additionally, it's important to note that this analysis only considers i.i.d. data among data-owners and a shallow DNN classifier, potentially masking further complications with main task convergence and the traitor tracing compatibility of black-box trigger features.

\bibliographystyle{ieeetr}
\bibliography{main}

\end{document}